\documentclass[12pt]{article}

\usepackage{epsf}
\usepackage[dvips]{graphicx}

\let\ensm=\ensuremath
\def\Pp{{\rm p}}

\def\PK{{\rm K}}

\def\pipm              {\ensm{\pi^\pm}}

\def\gevp             {\ensm{\,{\rm GeV}/c}}

\begin{document}

\begin{center}
{\bfseries 
SINGLE SPIN ASYMMETRY OF CHARGED HADRON PRODUCTION BY 40 GEV/C
 POLARIZED PROTONS
}

\vskip 5mm

 \underline{V.V.~Abramov}$^{1 \dag}$, P.I.~Goncharov$^{1}$,
 A.Yu.~Kalinin$^{1}$,
 A.V.~Khmelnikov$^{1}$, A.V.~Korablev$^{1}$,  Yu.P.~Korneev$^{1}$,
 A.V.~Kostritsky$^{1}$, A.N.~Krinitsyn$^{1}$, V.I.~Kryshkin$^{1}$,
 A.A.~Markov$^{1}$,  V.V.~Talov$^{1}$,  L.K.~Turchanovich$^{1}$, 
  A.A.~Volkov$^{1}$

\vskip 5mm

{\small
(1) {\it
Institute for High Energy Physics, Protvino, Moscow region, Russia
}
\\
$\dag$ {\it
E-mail: Victor.Abramov@ihep.ru
}}
\end{center}

\vskip 5mm

\begin{center}
\begin{minipage}{150mm}
\centerline{\bf Abstract}
The single spin asymmetry for charge hadron production off
 nuclei (C, Cu) has been measured using 40 GeV/c polarized proton beam.
 The measurements were carried out using FODS-2 experimental setup at
 IHEP. The data are presented in the central region and the forward
 region with respect to the incident protons.
\end{minipage}
\end{center}
\section{Introduction}

 The  single spin asymmetries (SSA) 
in hadron-hadron interactions are much larger than expected from 
the naive pertubative QCD \cite{Kane}. Also, its dependence on kinematic
 variables and hadron type
is far from the 
complete understanding. The existing data are measured in 
a limited range of c.m. production angles, transverse momenta and energies.
The previous measurements of SSA have shown that its absolute value is
rising with the increase of $p_T$ and  $x_F$ for beam energies above 20 GeV
\cite{E704pip},\cite{E925}. 
There were no measurements so far of the single spin asymmetries of the charged
hadron production in the energy range between 22 and 200 GeV. 
We have measured the single-spin asymmetry $A_N$ of the inclusive charged
pion, kaon, proton and antiproton production cross sections at high
$p_T$ and high $x_F$ regions for a 40 $\gevp$ polarized proton beam incident 
on nuclei (C, Cu), where $A_N$ is defined as
 \begin{equation}
 A_{N} = { 1\over{P_{B}\cdot cos{\phi}} } \cdot
{ {N{\uparrow} - N{\downarrow}} \over {N{\uparrow} + N{\downarrow}} }, \quad
\end{equation}
where $P_B$ is the beam polarization, $\phi$ is the azimuthal angle of the 
production plane, ${N\!\!\uparrow}$ and ${N\!\!\downarrow}$ are event rates
for the beam spin up and down respectively. The measurements were
carried out at IHEP, Protvino in 2003. The preliminary results in the central
 region were reported elsewhere \cite{ITEP04}.
The measurements of  $A_N$ in the central region using hydrogen target
 revealed a sizable asymmetry for $\pi^{+}$ and K$^{+}$ production 
\cite{NP492}. 
\section{Polarized beam and experimental setup}
The polarized protons are produced by the parity - nonconserving 
$\ensm{\Lambda}$ decays \cite{ITEP04},\cite{NP492}.
 The up or down beam transverse
polarization is achieved by the selection of decay protons with angles
near $90^{\circ}$ in the $\ensm{\Lambda}$ rest frame by
a movable collimator.  At the end of the beam line two  magnets 
 correct the vertical beam position on the spectrometer target for the two
beam polarizations. The intensity of the 40 $\gevp$ momentum
polarized beam on the spectrometer target  is $3\times 10^{7}$ ppp,
 $\Delta p/p= \pm 4.5$\%, the transverse polarization is $39^{+1}_{-3}$~\%,
 and the polarization direction
is changed each 18 min during 30 s. The beam intensity and the position
are measured by the  ionization chambers and the scintillation
hodoscopes.

\begin{figure}[h]
 \centerline{
 \includegraphics[width=90mm,height=70mm]{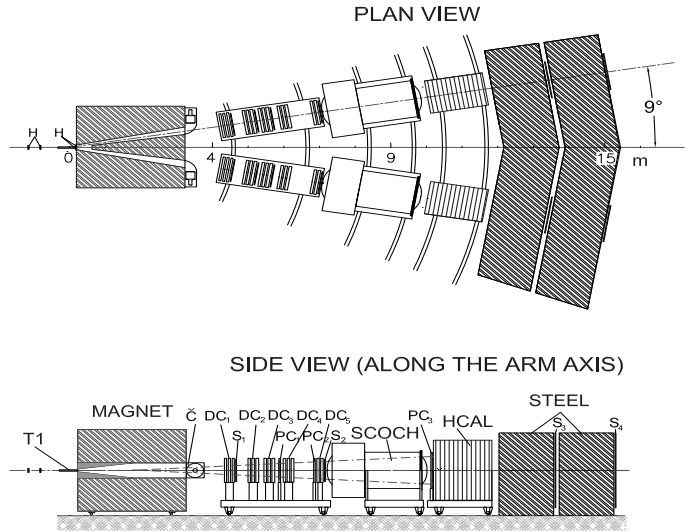}
}
 \caption{Schematic of experimental layout FODS-2.
 \label{fods}}
\end{figure}

  The measurements were carried out with the FODS-2 
 \cite{ITEP04},\cite{NP492} spectrometer (Fig.~\ref{fods}).
  It consists of an analyzing magnet, the drift chambers, the
  Cherenkov radiation spectrometer (SCOCH) for the particle
identification ($\pipm$ , $\PK^{\pm}$ , $\Pp$ and $\bar{\rm{p}}$), 
 the scintillation counters, and the
hadron calorimeters to  trigger on the high energy hadrons. 
 Inside the magnet there is also a beam dump made of tungsten
and copper.  There are two arms which can be rotated around the target
center situated in front of the magnet to change the secondary
particle angle.

 There are two threshold Cherenkov
counters using air at the atmospheric pressure inserted in the
magnet which are used for further improvement of particle identification.  
\section{Measurements}
 The measurements of $A_{N}$ in
the range $-0.15 \le x_{F} \le 0.2$ and $0.5 \le p_{T} \le 4$ GeV/$c$ 
are carried out with the symmetrical
arm positions at  angles of $\pm$ 160 mrad 
(mean c.m. angle $\theta_{cm} = \pm$ 86$^o$).
 The results for the two arms and
  the different values of the magnetic field B in the spectrometer are
averaged, which partially cancels systematical uncertainties connected with the
variation of the beam position in the vertical direction, the
intensity monitor and the apparatus  drift. Another set of measurements is
 carried out with the arms, rotated by 70 mrad 
(the left arm is at $\theta_{cm} = 48^o$ 
with $0.05 \le x_{F} \le 0.7$ and $0.5 \le p_{T} \le 2.5$ GeV/$c$ , 
and the right arm is at $\theta_{cm} = 105^o$  
with $-0.25 \le x_{F} \le -0.05$ and $0.7 \le p_{T} \le 3$ GeV/$c$ ). 
Both polarities of magnetic field B in the spectrometer magnet
 are used for data taking to reduce systematic uncertainties.
 In addition, two values of the magnetic field
 (B and B/2) are used  to extend the momentum range of the data. 
\section{Data processing}
 The reconstructed trajectory of a particle downstream the spectrometer 
magnet and beam coordinates from the beam hodoscopes are used to determine
 its  momentum, production angles and vertex $Z$-coordinate. 
After all cuts are applied, the remaining events are identified in the SCOCH
 and threshold cherenkov counters. The integrated beam flux is measured by 
the ionization chamber and is used to calculate normalized particle 
rates ($N{\uparrow}$ and $N{\downarrow}$ ) for two signs of the beam
 polarization. The beam coordinates, measured by the $X$ and $Y$
 hodoscope planes,
are used to estimate the mean beam position ($X_{B}$ and $Y_{B}$) 
during a spill, separately for ``UP'' and ``Down'' polarization signs.
 It was found that the mean beam coordinates, averaged over a group of runs
 with  similar conditions, can differ for ``UP'' and ``Down'' beam
 polarizations up to $\pm 0.5$ mm. Since the normalized rates
$N{\uparrow}$ and $N{\downarrow}$ depend on the beam position, 
the false asymmetry can be added to the analyzing power $A_{N}$. Cuts on 
beam coordinates are imposed to level the $X_{B}$ and $Y_{B}$   
for  ``UP'' and ``Down'' polarization sign with 4 ${\mu}m$ accuracy
 to avoid the false asymmetry. The other sources of systematic  
uncertainties, remaining after the above cuts are applied, 
contribute up to 4\% to the systematic error $\epsilon$, which is
 estimated from run to run  $A_N$ variation. The $\epsilon$ is added
 in the quadrature to the statistical error at each data point.
\section{Single spin asymmetries}
\begin{figure}[htb]
\centering
\begin{tabular}{cc}
\begin{minipage}{65mm}
\includegraphics[width=60mm,height=65mm]{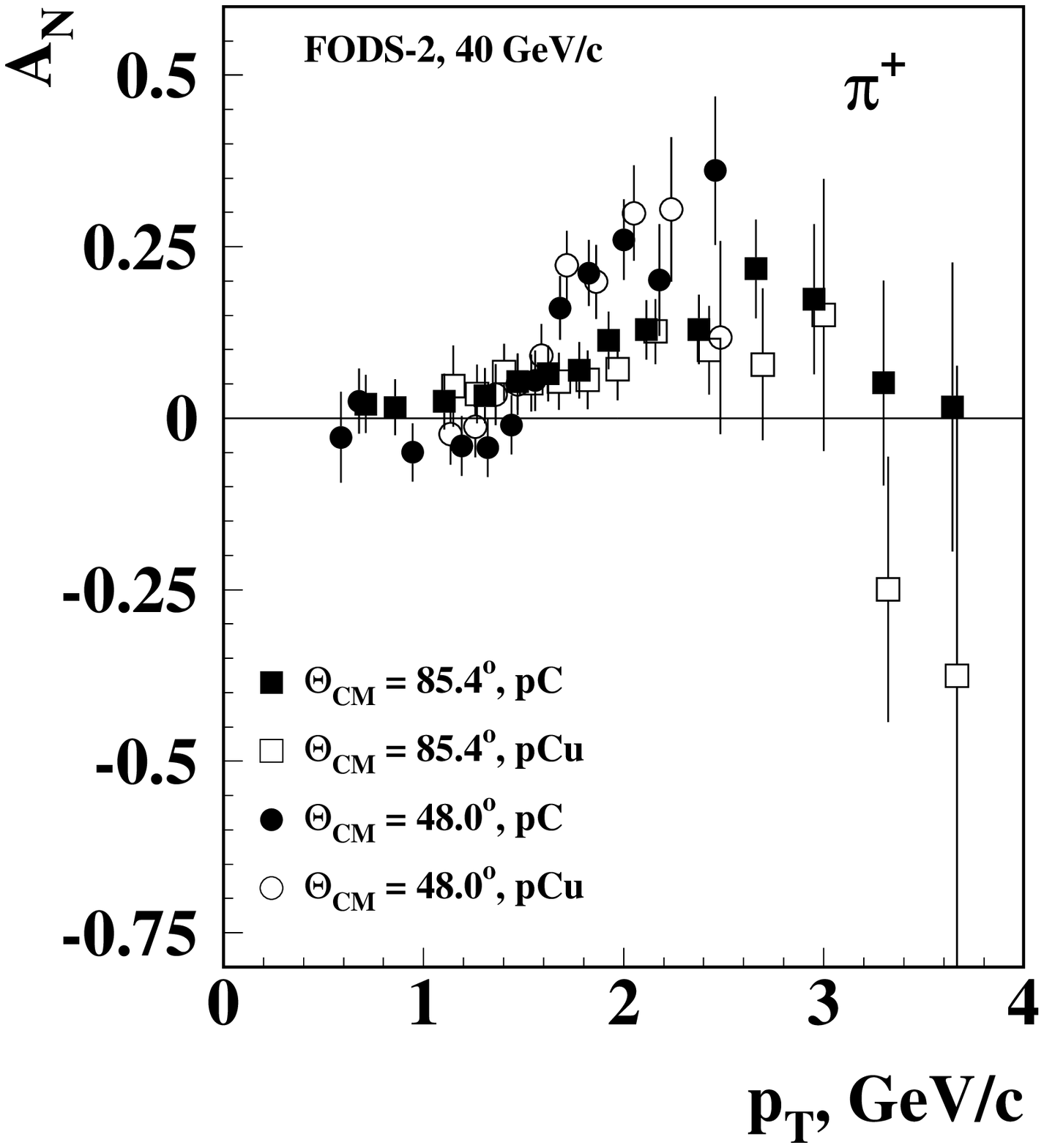}
\caption{ $A_{N}$ dependence
 on  $p_T$ for $\rm{p\! \uparrow + C(Cu) \rightarrow \pi^{+} + X}$. 
\label{pip}} 
\end{minipage}
& 
\begin{minipage}{65mm}
\includegraphics[width=60mm,height=65mm]{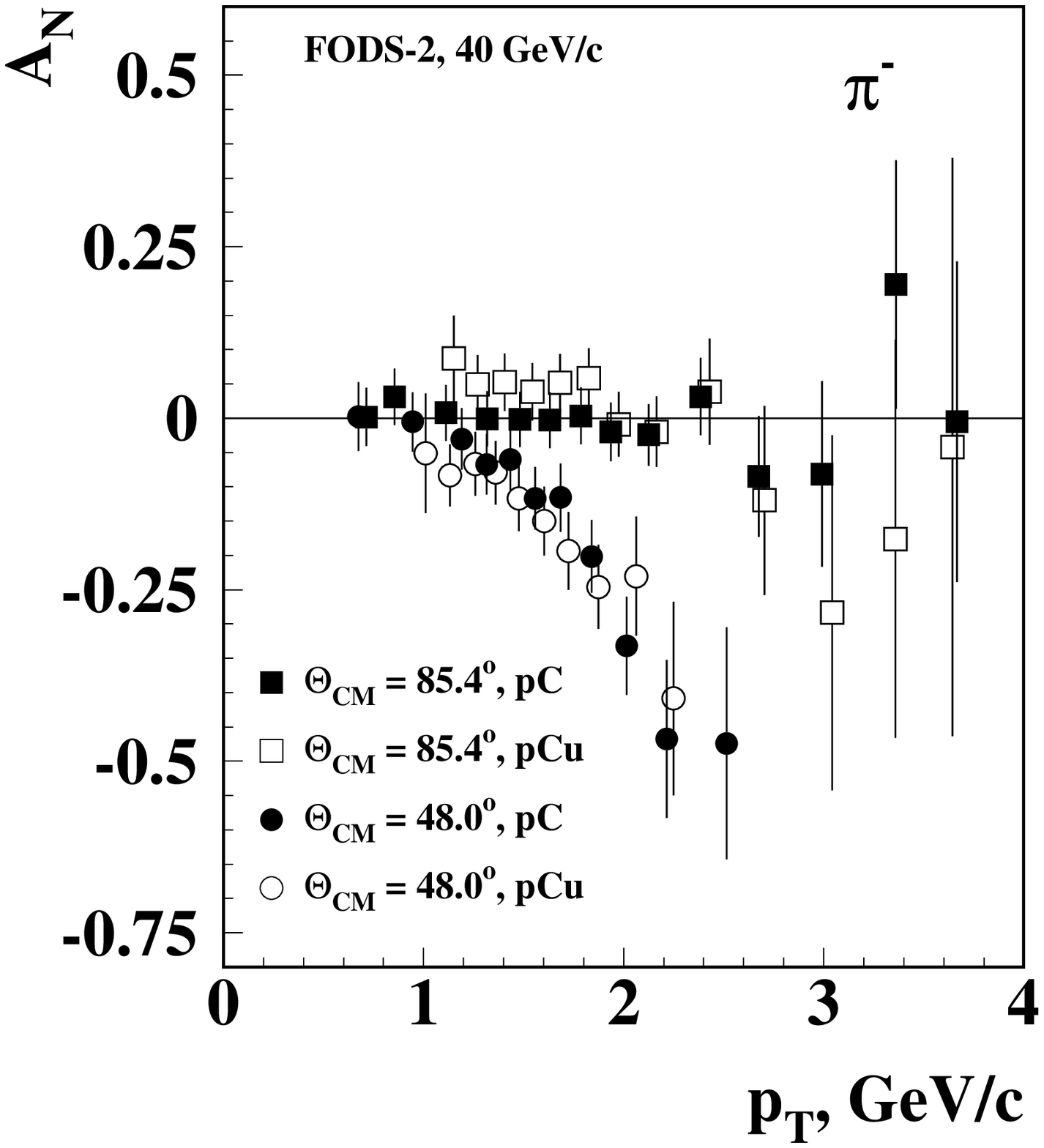}
\caption{ $A_{N}$ dependence
 on  $p_T$ for $\rm{p\! \uparrow + C(Cu) \rightarrow \pi^{-} + X}$. 
\label{pim}} 
\end{minipage} \\
\end{tabular}
\end{figure}

\begin{figure}[htb]
\centering
\begin{tabular}{cc}
\begin{minipage}{65mm}
\vskip -4mm
\includegraphics[width=60mm,height=65mm]{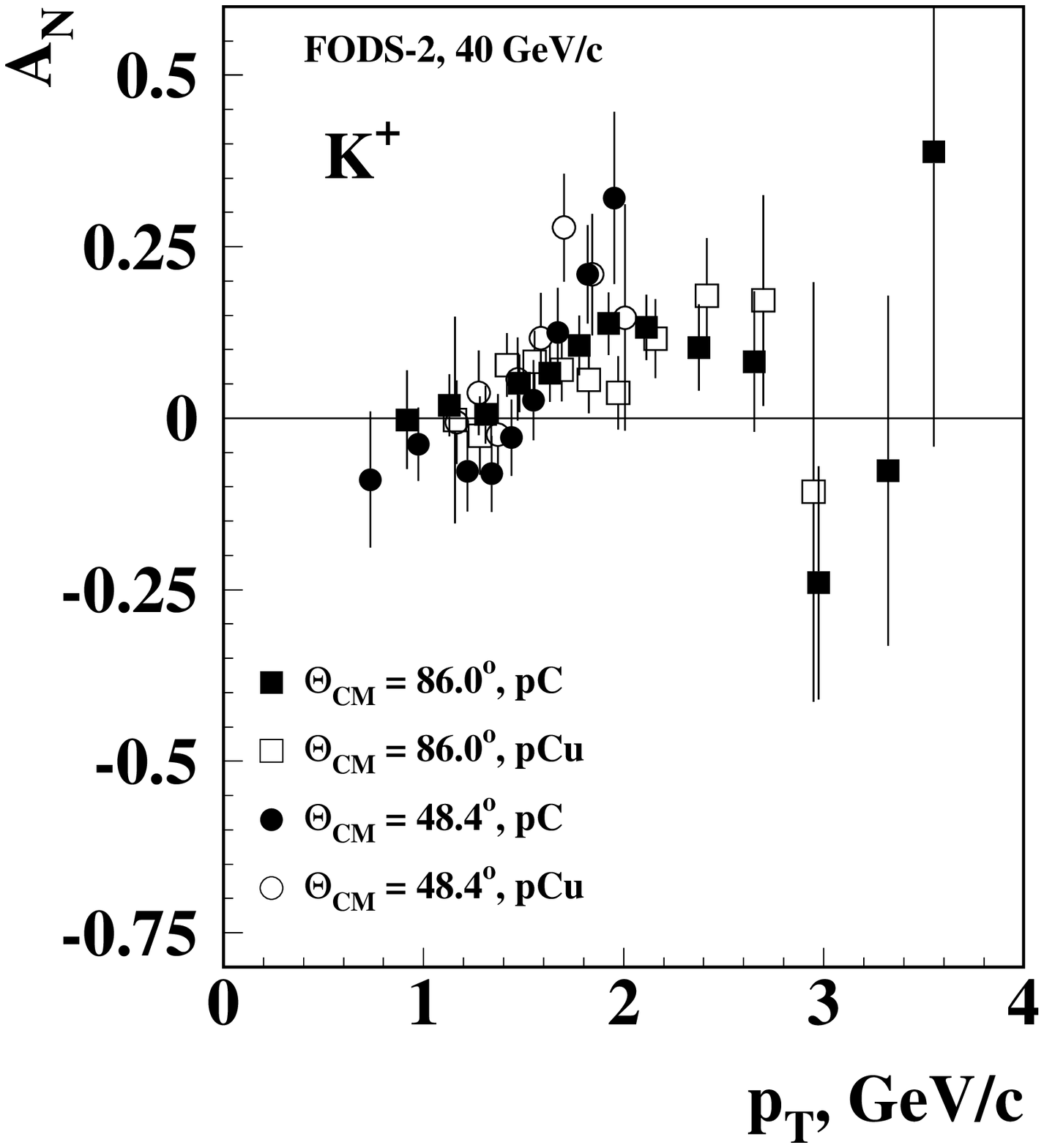}
\caption{ $A_{N}$ dependence
 on  $p_T$ for $\rm{p\! \uparrow + C(Cu) \rightarrow K^{+} + X}$. 
\label{Kp}} 
\end{minipage}
& 
\begin{minipage}{65mm}
\vskip -4mm
\includegraphics[width=60mm,height=65mm]{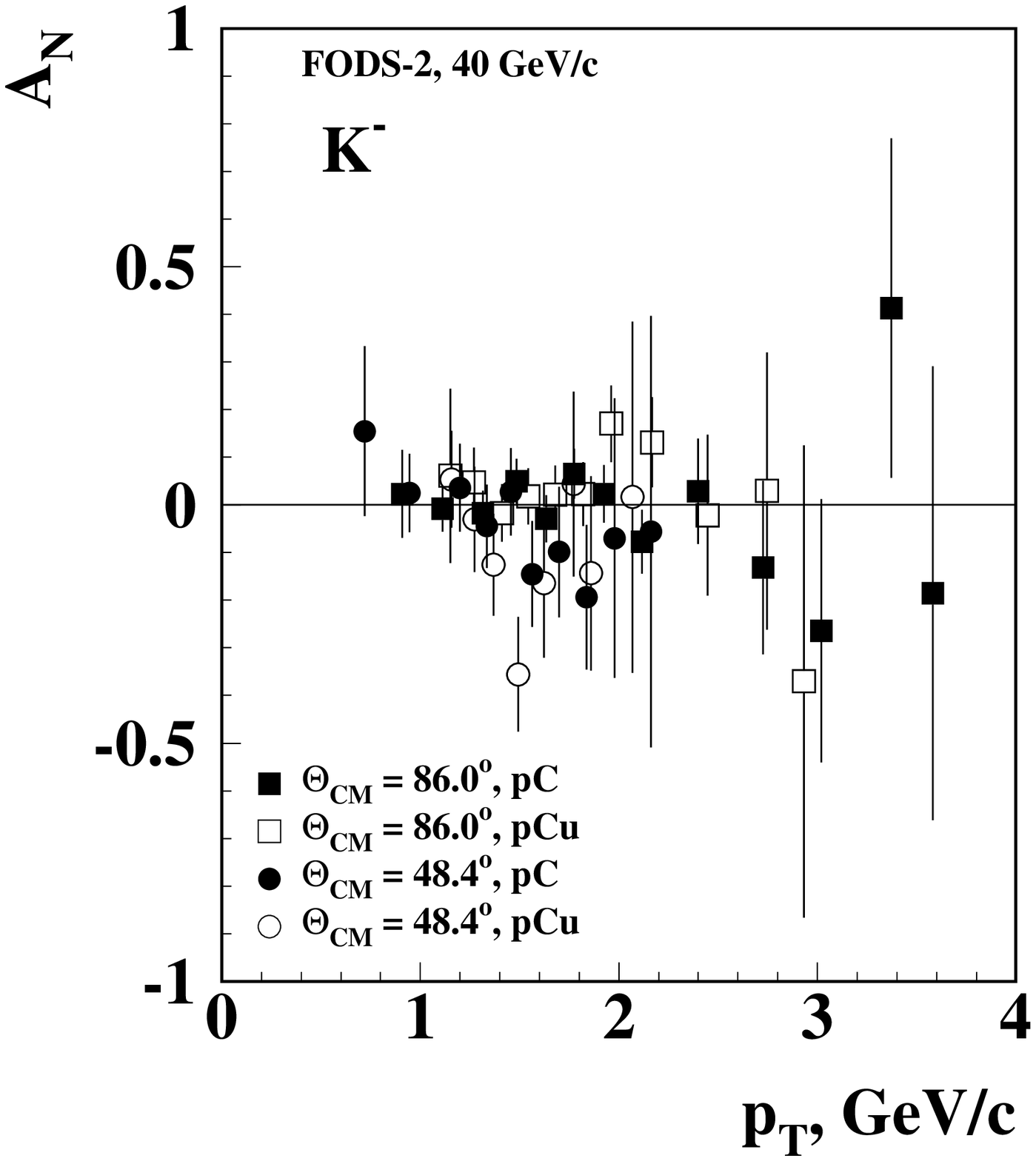}
\caption{ $A_{N}$ dependence
 on  $p_T$ for $\rm{p\! \uparrow + C(Cu) \rightarrow K^{-} + X}$. 
\label{Km}} 
\end{minipage} \\
\end{tabular}
\end{figure}

\begin{figure}[htb]
\centering
\begin{tabular}{cc}
\begin{minipage}{65mm}
\vskip -10mm
\includegraphics[width=60mm,height=65mm]{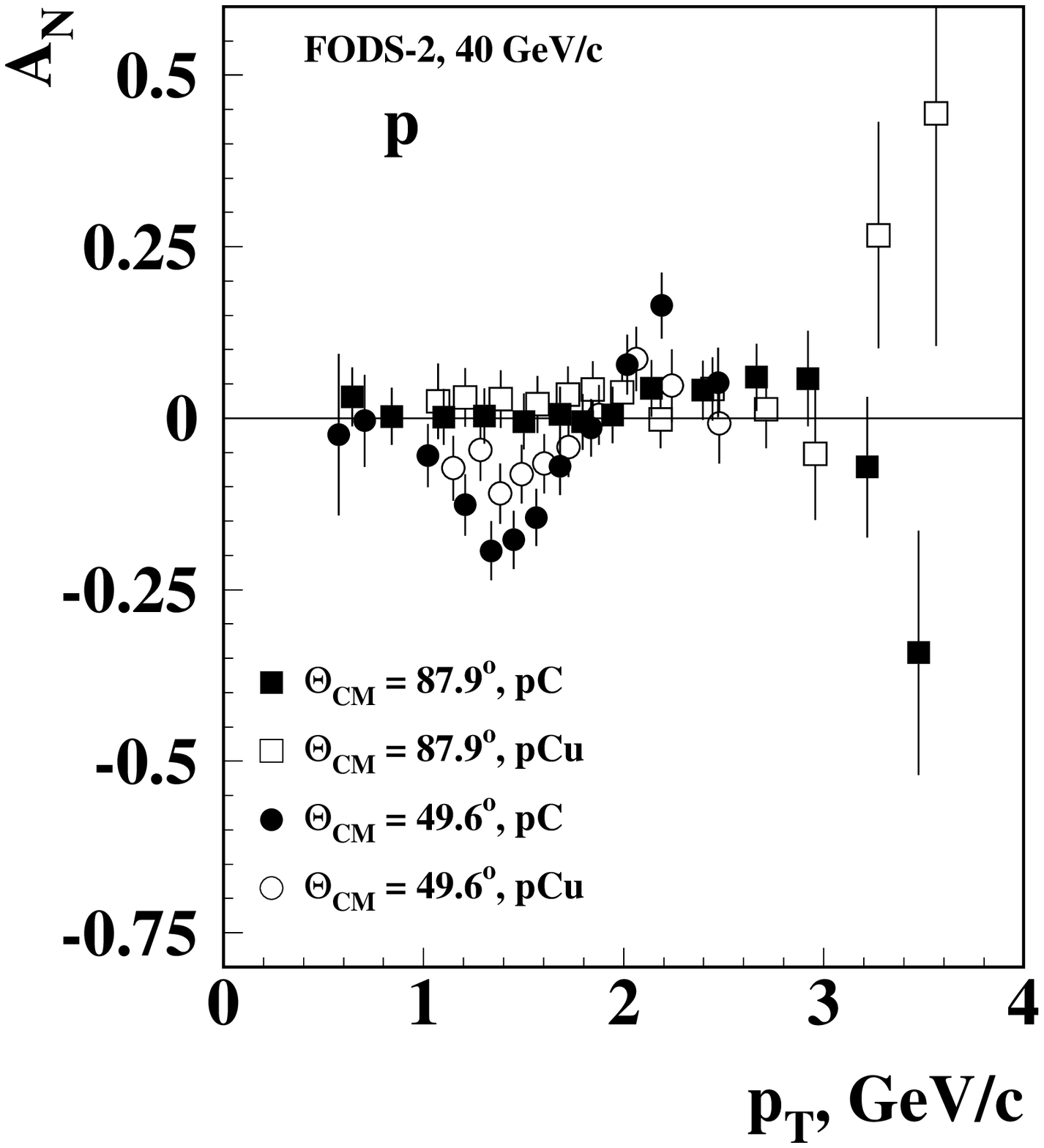}
\caption{ $A_{N}$ dependence
 on  $p_T$ for $\rm{p\! \uparrow + C(Cu) \rightarrow p + X}$. 
\label{pp}} 
\end{minipage}
& 
\begin{minipage}{65mm}
\vskip -10mm
\includegraphics[width=60mm,height=65mm]{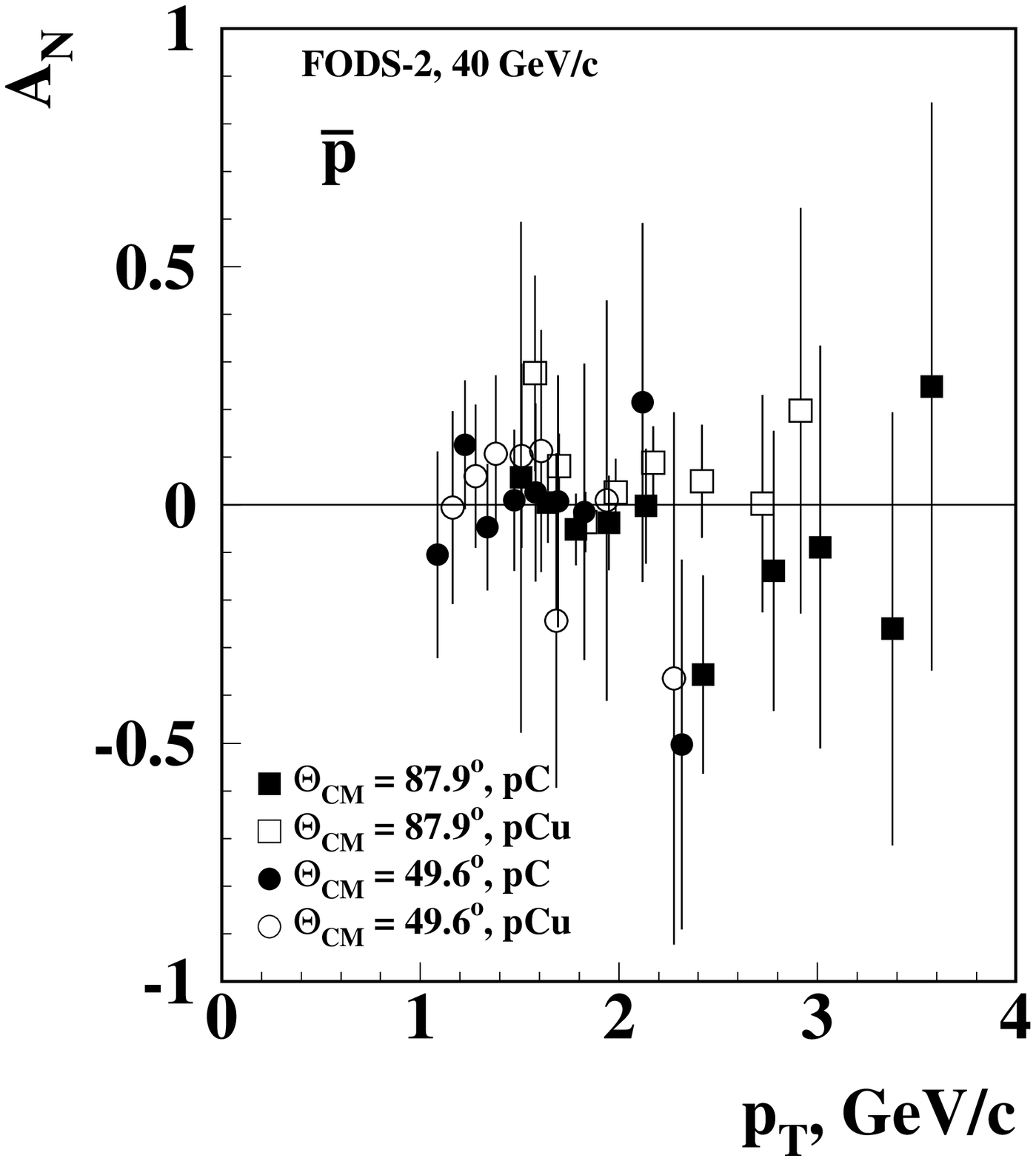}
\caption{ $A_{N}$ dependence
 on  $p_T$ for $\rm{p\! \uparrow + C(Cu) \rightarrow \bar{p} + X}$. 
\label{pm}} 
\end{minipage} \\
\end{tabular}
\end{figure}

\begin{figure}[htb]
\centering
\begin{tabular}{cc}
\begin{minipage}{65mm}
\vskip -4mm
\includegraphics[width=60mm,height=65mm]{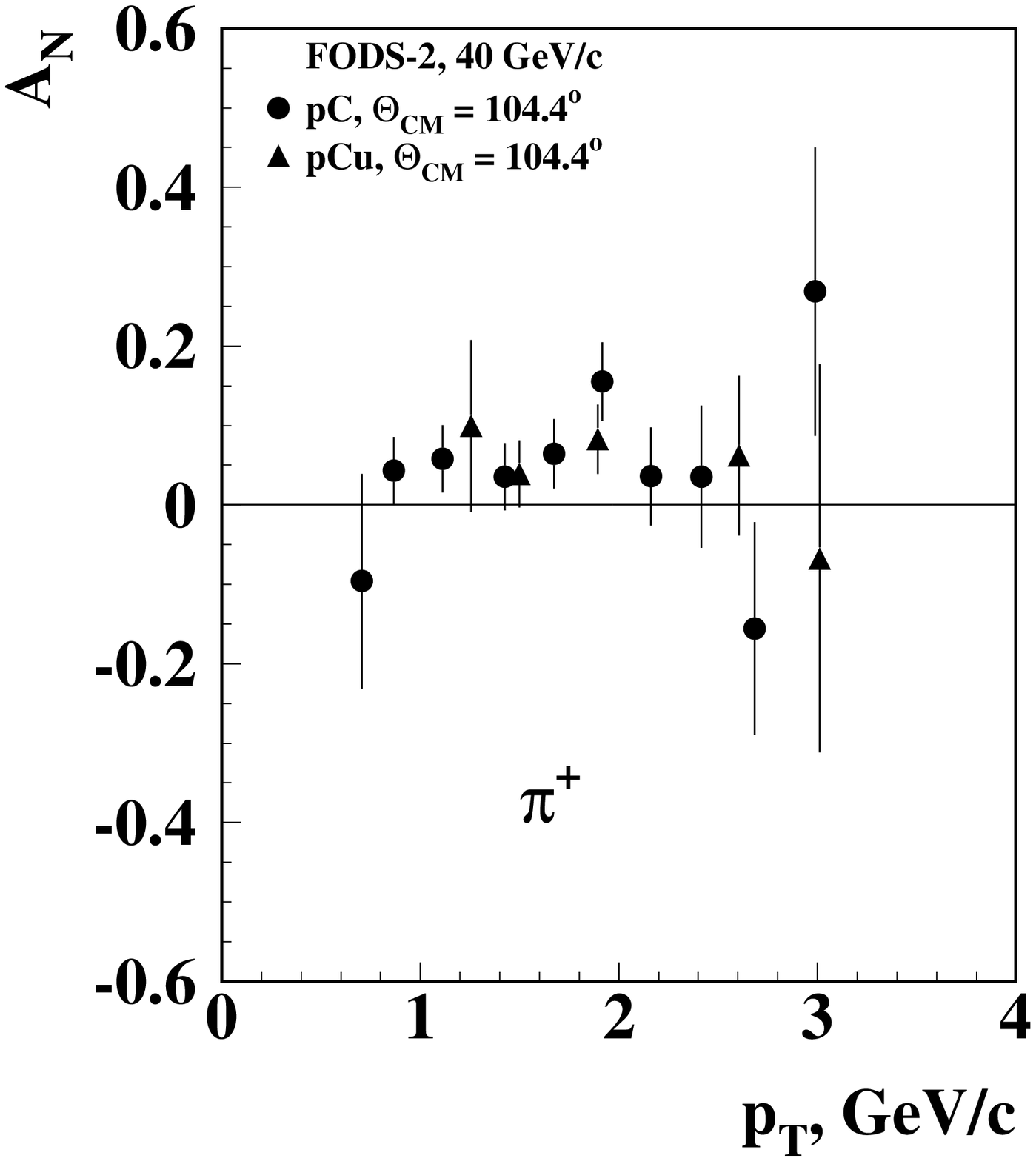}
\vskip -2mm
\caption{ $A_{N}$ vs   $p_T$ at $104.4^o$. 
\label{pip105}} 
\end{minipage}
& 
\begin{minipage}{65mm}
\vskip -4mm
\includegraphics[width=60mm,height=65mm]{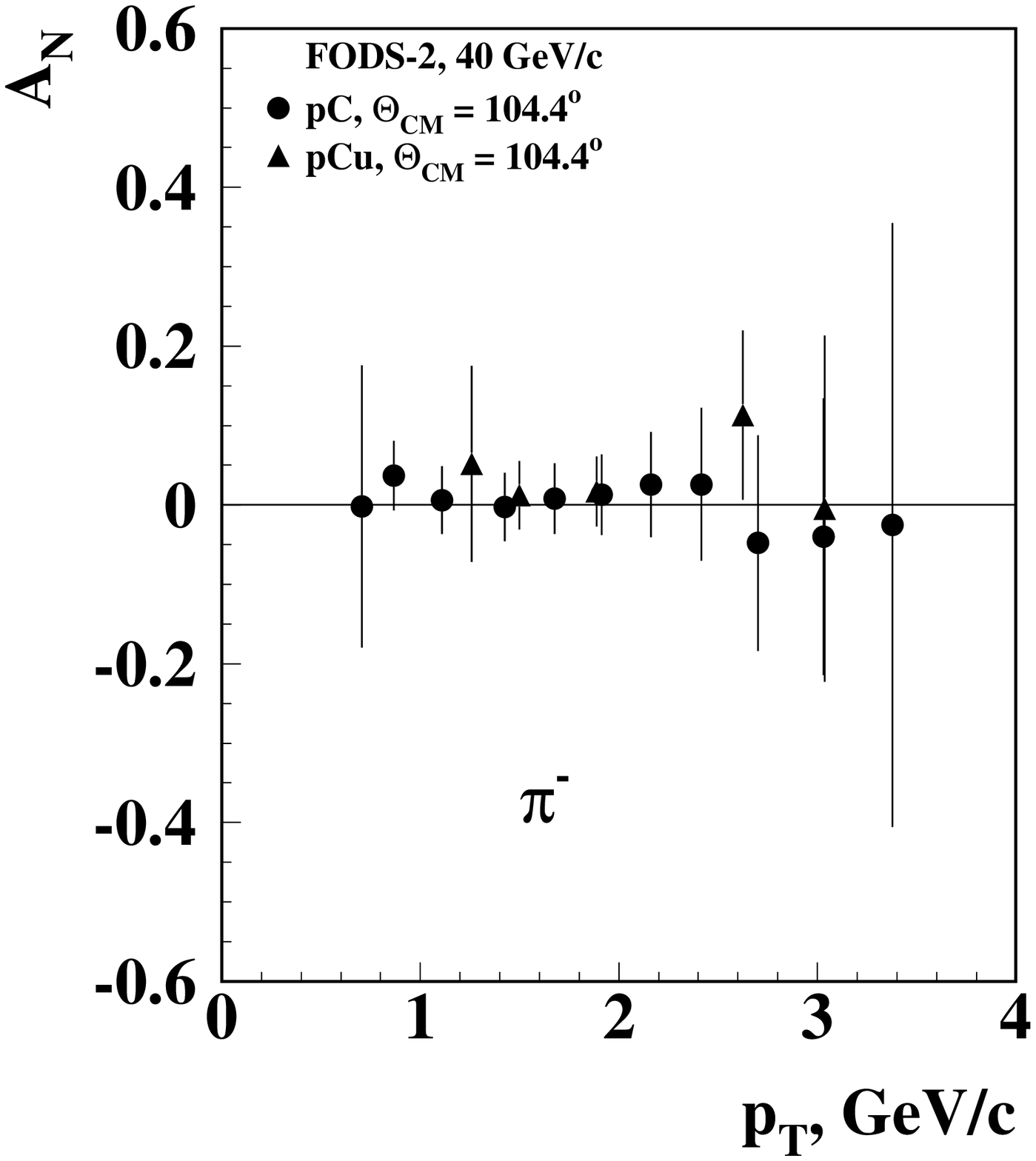}
\vskip -2mm
\caption{ $A_{N}$ vs   $p_T$ at $104.4^o$. 
\label{pim105}} 
\end{minipage} \\
\end{tabular}
\end{figure}

  The dependence of $A_N$ on transverse momentum ($p_T$) for $\pipm$ , 
$\PK^{\pm}$ , $\Pp$ and $\bar{\rm{p}}$ production on C and Cu targets and
at two c.m. angles ($\approx 48^o$ and $\approx 85^o$)  are shown
in Figs.~\ref{pip} -~\ref{pm}. 
The significant value of $A_N$ is seen in $\pi^{+}$,  $\pi^{-}$, $K^{+}$ 
and proton production at $\theta_{cm} \approx 48^o$ on both targets. 
The $A_N$ at  $\theta_{cm} \approx 85^o$ is approximately two times smaller in 
$\pi^{+}$ and $K^{+}$ production and reveals a breakdown 
at $p_T\approx 2.5$ GeV/c in its $p_T$ dependence, which could indicate
 a transition to the predicted pQCD regime \cite{Kane}. The $A_N$ for $K^{-}$
and $\bar{p}$ production is consistent with zero for both targets 
and all c.m. angles, as expected due to small sea quark polarization. 
The proton data at  $\theta_{cm} \approx 48^o$ reveal an oscillation of $A_N$ 
with minimum at 1.3 GeV/c and maximum near 2.2 GeV/c.
The SSA at  $\theta_{cm} \approx 105^o$ are shown
 in Figs.~\ref{pip105} and ~\ref{pim105} for $\pi^{+}$ and  $\pi^{-}$,
 respectively. For all charged hadrons the SSA is close to zero at
 $\theta_{cm}$ near $105^o$. No significant A-dependence is observed 
in the above data.

  Comparison of FODS-2 results with the data, measured at 22 GeV/c \cite{E925}
and at 200 GeV/c \cite{E704pip}, is shown in Figs.~\ref{pip105} -~\ref{pim105} 
as a function of a scaling variable $X_{S} = (X_{R}+X_{F})/2 -X_{0}$, where
$X_{0} = 0.075N_{Q} + 2N_{Q}M_{Q}(1+\cos\theta_{cm})/\sqrt{s}$
takes into account the  constituent quark mass $M_Q=0.3$ GeV/c$^2$ and
the number $N_Q$ of valence quarks in the observed hadron.
For all three energies the $\pi^+$ SSA in the forward region is described
 well by a single function of $X_S$. The $\pi^-$ and proton SSA
agree with E925 data for $p_T \ge 0.6$ GeV/c \cite{ITEP04}. References to 
other examples of scaling SSA behavior can be found in Ref. \cite{ITEP04}.     

  In conclusion, the SSA are measured on C and Cu targets 
for  $\pi^{\pm}$, $K^{\pm}$,
 $\bar{p}$ and proton production
 at mean c.m. angles $48^o$, $86^o$ and $105^o$.
The decrease of SSA above 2.5 GeV/c for  $\pi^{+}$, $K^{+}$ and protons
can indicate a transition to the pQCD regime, where $A_N$ tends to zero. 
No significant A-dependence is observed for the SSA. The SSA for  $K^{-}$
and $\bar{p}$ are consistent with zero, as expected due to the small sea
quark polarization. The scaling behavior of SSA is seen
 for $\theta_{cm} \le 50^{o}$ and $p_{T} \ge 0.6$ GeV/c. 
The SSA is close to zero for  $\theta_{cm} \approx 105^{o}$.

We are grateful to the IHEP staff for their assistance with setting up
the experiment and the IHEP directorate for their support.
\begin{figure}[htb]
\centering
\begin{tabular}{cc}
\begin{minipage}{65mm}
\vskip -8mm
\includegraphics[width=60mm,height=65mm]{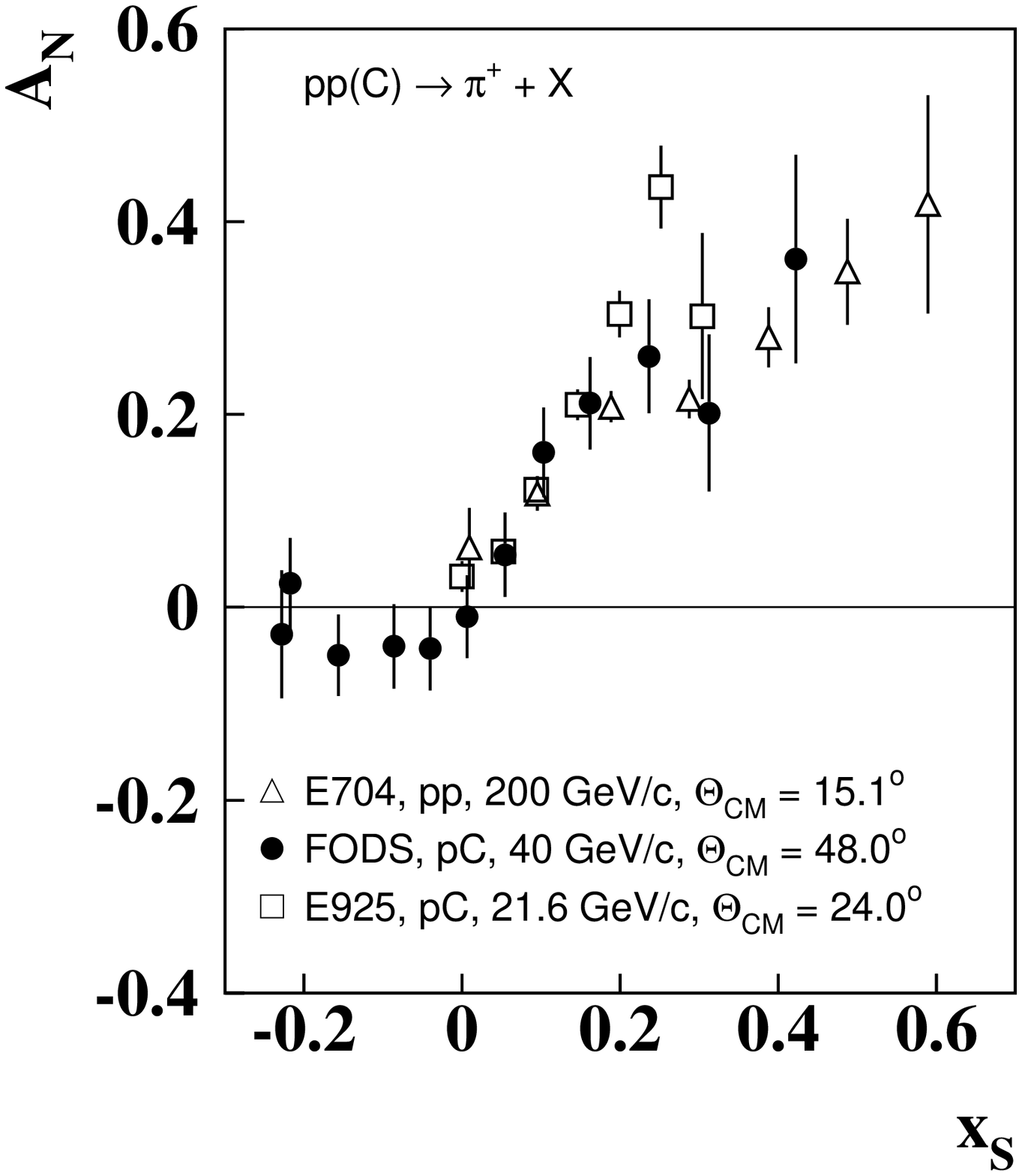}
\vskip -4mm
\caption{ $A_{N}$ vs $X_S$ for $\pi^{+}$ production at 22,
 40, and 200 GeV. 
\label{pips}} 
\end{minipage}
& 
\begin{minipage}{65mm}
\vskip -8mm
\includegraphics[width=60mm,height=65mm]{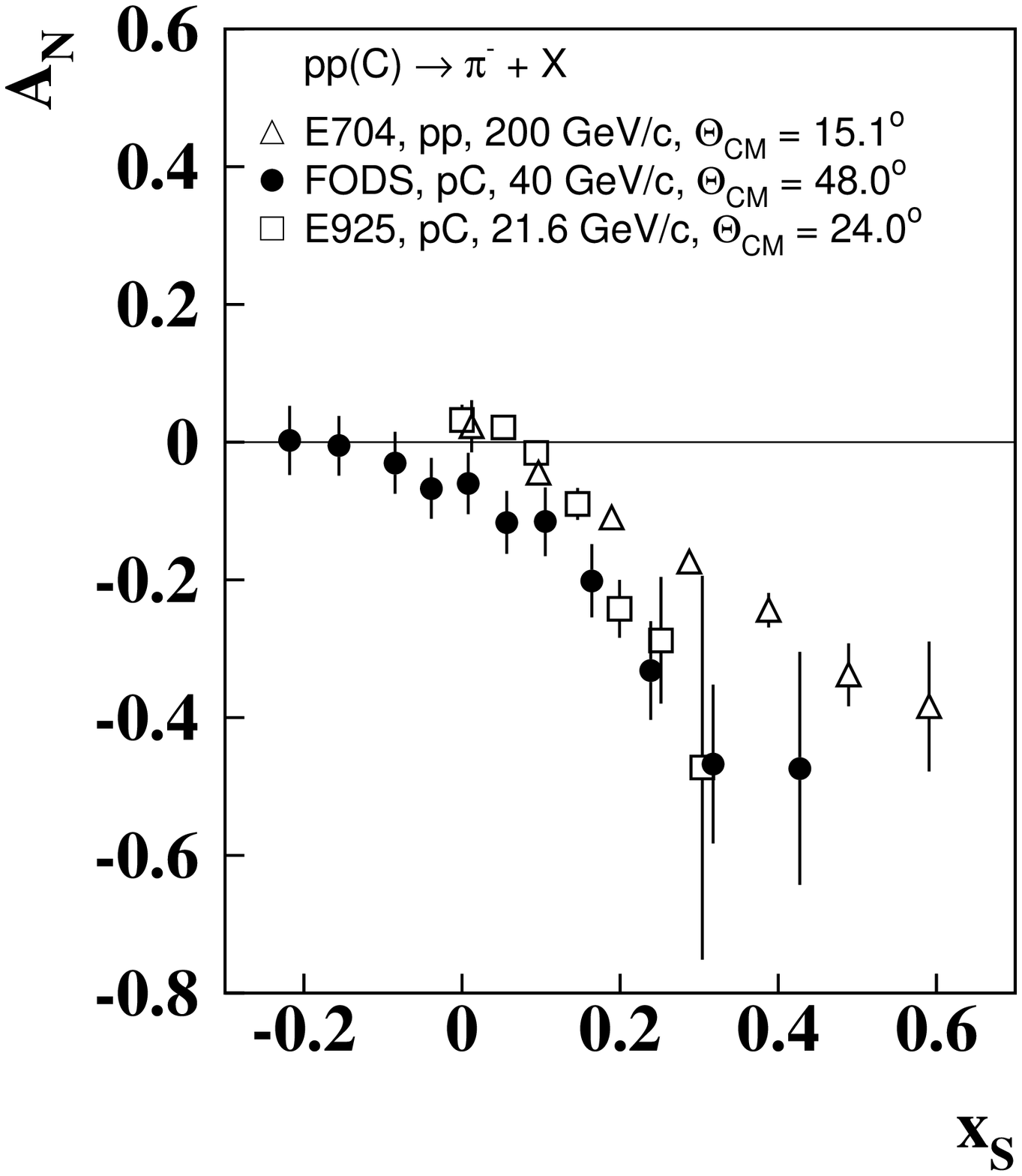}
\vskip -4mm
\caption{ $A_{N}$ vs $X_S$ for $\pi^{-}$ production at 22,
 40, and 200 GeV. 
\label{pims}} 
\end{minipage} \\
\end{tabular}
\end{figure} 

\end{document}